\documentclass[11pt]{article}
\usepackage{moriond,epsfig}

\def\Journal#1#2#3#4{{#1} {\bf #2}, #3 (#4)}

\def\PRD{{\em Phys. Rev.} D}

\def\be{\begin{equation}}
\def\ee{\end{equation}}
\def\bea{\begin{eqnarray}}
\def\eea{\end{eqnarray}}

\begin{document}
\vspace*{4cm}
\title{ELECTRICAL CONDUCTIVITY OF QUARK MATTER\\
 IN MAGNETIC FIELD~\footnote{Contribution to the Proceedings of the XLVI-th Recontres de Moriond QCD and Hadronic Interactions, LaThuile, Italy, march 2011}}

\author{ B. KERBIKOV }

\address{ITEP, Moscow, Russia}

\author{M. ANDREICHIKOV}

\address{MIPT--NBIC, Moscow, Russia}

\maketitle\abstracts{Fermion currents in dense quark matter embedded into magnetic field are under intense discussions motivated by Chiral Magnetic Effect. We argue that conductivity of quark matter may be independent of the magnetic field direction and not proportional to the magnetic field strength.}


Magnetic field created by heavy ion currents at RHIC and LHC at the collision moment is huge, $ |eB| \geq m_{\pi}^2 \sim 10^{18} \ \mathrm{G} $ ~\cite{work1}. An intriguing effect observed by STAR collaboration and first reported at the XLIV-th Recontres de Moriond QCD~\cite{work2} in the electric current induced in the direction of the magnetic field --- Chiral Magnetic Effect~\cite{work3}. The relaxation time of the magnetic field crucially depends on the quark matter electric conductivity~\cite{work4}. It is clear that the problem of quark matter conductivity in magnetic field is an important albeit a complicated one. The approach to the problem which we briefly present below patterns on the method developed in condensed matter physics~\cite{work5}. Kubo formula relates the conductivity to a two--point correlator of the current:
\begin{equation}
\sigma_{lm}(i\omega_k,\mathbf{q}) = \frac{e^2 T}{\omega_k}  \mathrm{Tr} \sum_{\mathcal{E},\mathbf{p}} G^M(\mathbf{p}, \tilde{\mathcal{E}}_n) \gamma^l G^M(\mathbf{p} + \mathbf{q}, \tilde{\mathcal{E}}_n + \omega_k) \gamma^m,
\label{equ1}
\end{equation} 
where $ \mathrm{Tr} $ is taken over Dirac indices, color and flavour indices are omitted, $ G^M $ is the relativistic Matsubara propagator, $ \tilde{\mathcal{E}}_n = \mathcal{E}_n + \frac{1}{2\tau}sgn(\mathcal{E}_n) $, $ \mathcal{E}_n = \pi T (2n+1) $, $ \tau $ is the momentum relaxation time. In the disordered system quark acquires the self--energy proportional to the inverse relaxation time $ \tau $ on chaotically distributed scatterers. Depending on the averaging procedure of the correlator (\ref{equ1}) over the disorder one obtains two different sets of diagrams giving two contributions to the conductivity, namely the Drude (Boltzmann) one $ \sigma_{cl} $ and the so--called quantum correction $ \sigma_q $. Relation between the two is given by $ \sigma_q \simeq (p_Fl)^{-1} \sigma_{cl} $, where $ p_F $ is the Fermi momentum and $ l $ is the quark mean free path. The values of these parameters depend upon the location of the system on the QCD phase diagram in the $ (T, \mu) $ coordinates. As an example, we take the quark chemical potential $ \mu \simeq 0.4 \ \mathrm{GeV} $, consider chiral quarks $ p_F \simeq \mu $, and $ l \simeq 0.5 \div 1 \ \mathrm{fm} $. Then $ \sigma_q \geq 0.5 \sigma_{cl} $ and the term ``correction'' used in condensed matter physics is no more meaningful. Omitting the derivation we present the resulting expression for the frequency dependent conductivity:
\begin{equation}
\sigma(\omega) = \sigma_{cl} + \sigma_{q} = \frac{ne^2}{m} \frac{\tau}{1+\omega \tau} - \frac{2 \mathcal{D}e^2}{\pi} \int \! \frac{\mathrm{d}^3 \mathbf{q}}{(2\pi)^3} \frac{1}{-i\omega + \mathcal{D} \mathbf{q}^2}
\label{equ2}
\end{equation}
Here $ \mathcal{D} $ is the diffusion coefficient and the appearance of the slow diffusion mode is an important feature of the quantum conductivity. The first term is the Drude conductivity. Similar structure of the conductivity emerges in the hydrodynamic approach to strongly coupled CFT~\cite{work6}. Our interest here is $ \sigma_q $. The negative sign reflects the fact that due to quantum interference the probability of quark returns increases, at $ (p_F l) > 1 $ the system undergoes Anderson transition and becomes an insulator~\cite{work7}. One may view quantum conductivity as being originated by the presence of a fictitious spin--zero particle with a charge $ 2e $ and a mass $1 / 2\mathcal{D} $. The simplest model of this particle would be a fluctuating Cooper pair. The fact that the effective charge carrier is a scalar particle is at the core of the unusual behaviour of quantum conductivity in magnetic field. To introduce the magnetic field we choose the gauge $ A_x = 0 $, $ A_y = Bx $, $ A_z = 0 $, so that $ \mathbf{B} $ is directed along the $z$--axis. There are three characteristic length scales in the problem: the mean free path $ l $, the magnetic length $ l_B = (eB)^{-1/2} $, and the phase--randomizing length $ l_{\varphi} = (2\mathcal{D}\tau_{\varphi})^{1/2} $, where $ \tau_{\varphi} $ is the phase--breaking time due to inelastic processes. We assume that $ l_{\varphi} \gg l $ which is a questionable supposition for the quark matter. Returning to (\ref{equ2}), we may say that the characteristic momentum scale for $ \sigma_{cl} $ is $ p > 1/l $, while for $ \sigma_q $ it is $ p < 1/l $. In magnetic field and with phase--breaking interaction the denominator $ (-i\omega + \mathcal{D}\mathbf{q}^2) $ in (\ref{equ2}) is substituted by $\left[ -i\omega + \mathcal{D}q^2_z + \Omega(n+\frac{1}{2}) + \tau_{\varphi}^{-1} \right] $, with $ \Omega = 4eB\mathcal{D} $, $ n $ numerates Landau levels. Integrating over $ p_y $, making use of the completeness of the Landau wave functions and integrating over $ |p_z| < 1/l $, we obtain:
\begin{equation}
\sigma_q = -\frac{e^2}{\pi^3 l_B} \sum_{n=0}^{n_{\mathrm{max}}} \frac{1}{\sqrt{n+\frac{1}{2}+\delta}}\arctan \left( \frac{l_B}{2l\sqrt{n+\frac{1}{2}+\delta}} \right),
\label{equ3}
\end{equation}
where $ n_{\mathrm{max}} = l_B^2/l^2 $, $ \delta=l_B^2/l_{\varphi}^2 $. Truncation of the sum over the Landau levels at $ n_{\mathrm{max}} $ corresponds to the condition $ p < 1/l $ formulated above. When $ n_{\mathrm{max}} \gg 1 $ (weak field) we may substitute summation by integration and obtain:
\begin{equation}
\sigma_q = -\frac{e^2}{\pi^2}\left( \frac{1}{l} - \frac{1}{l_{\varphi}} \right).
\label{equ4}
\end{equation}
This means that for $ l_B \gg l_{\varphi} \gg l $ quantum conductivity does not feel the magnetic field. For $ l \simeq 1 \ \mathrm{fm} $, $ \tau_{\varphi} \sim 4\tau $ this corresponds to $ |eB| \ll  10^{4} \ \mathrm{MeV}^2 $, i.e. magnetic field at RHIC $ |eB| \sim m_{\pi}^2 \sim 2 \cdot 10^4 \ \mathrm{MeV}^2 $ is not too strong in the above sense.

Let us denote $ \sigma_q $ in this ``weak'' field limit by $ \sigma_q^{<} $ and for stronger field by $ \sigma_q^{>} $. It may be shown that $ |\sigma_q^{>}| < |\sigma_q^{<}| $ and
\begin{equation}
\sigma_q^{>} - \sigma_q^{<} \sim l_B^{-1} \sim \sqrt{eB}.
\label{equ5}
\end{equation}

Summarizing we may say that:
\begin{itemize}
\item[(i)]
quantum contribution is an important part of quark matter conductivity; 
\item[(ii)]
it makes the total conductivity smaller;
\item[(iii)]
it only weakly depends on the magnetic field and does not depend on the field direction.
\end{itemize}

Our final remark is that due to Lorentz contraction ultra--relativistic ions are effectively two--dimensional objects. In two--dimensional systems $ \sigma_q $ logarithmically diverges at $ \omega \to 0 $.

\section*{Acknowledgments}

One of the authors (B.K.) gratefully aknowledges support from the RFBR grants 10-0209311--NTSNIL-a, 11-02-08030-z and from the organizers of the Recontres de Moriond. This work was partly done during INT workshop ``Fermions from Cold Atoms to Neutron Stars''.

\end{document}